\documentclass[conference,10pt,a4paper]{IEEEtran}

\usepackage{silence}
\newlength{\flexwidth}
\usepackage{bm}
\usepackage{amsmath,amssymb,amsthm,fixmath}
\usepackage{mathtools}
\usepackage{accents}
\usepackage{color,colortbl}
\usepackage{xcolor}
\usepackage{siunitx}
\usepackage{cuted}
\usepackage{multirow,booktabs}
\usepackage{subcaption}
\usepackage[inline]{enumitem}
\usepackage{optidef}
\usepackage{graphicx}
\usepackage{lipsum}

\usepackage{amsmath}
\usepackage{epstopdf}
\usepackage[normalem]{ulem}

\usepackage[textsize=tiny,colorinlistoftodos]{todonotes}
\makeatletter
\define@key{todonotes}{bh}[]{% Comment by 
	\setkeys{todonotes}{author=\textbf{Bin}, color=lime!30}}%
\define@key{todonotes}{zf}[]{% Comment by 
	\setkeys{todonotes}{author=\textbf{Zexin}, color=blue!30}}%
\makeatother

\newif\ifreviewmode
\reviewmodetrue % Set to true to enable review mode
\reviewmodefalse

\ifreviewmode
\else
  \renewcommand{\todo}[1]{} % hide todo notes
\usepackage[shortcuts,acronym,automake]{glossaries}
\makeglossaries
\newacronym{ue}{UE}{User Equipment}
\newacronym{bs}{BS}{base station}
\newacronym{csi}{CSI}{Channel state information}
\newacronym{b5g}{B5G}{Beyond-Fifth-Generation}
\newacronym{6g}{6G}{Sixth Generation}
\newacronym{ml}{ML}{Machine learning}
\newacronym{sbs}{SBS}{small base station}
\newacronym{mu}{MU}{mobile user}
\newacronym{mbs}{MBS}{macro base station}
\newacronym{mse}{MSE}{Mean Squared Error}
\newacronym{cl}{CL}{centralized learning}
\newacronym{uav}{UAV}{Uncrewed Aerial Vehicle}
\newacronym{bme}{BME}{Bayesian Model Ensemble}
\newacronym{iid}{IID}{independent and identically distributed}
\newacronym{raf}{RAF}{robust aggregation function}
\newacronym{sgd}{SGD}{stochastic gradient descend}
\newacronym{cdf}{CDF}{cumulative distribution function}
\newacronym{lid}{LID}{local intrinsic dimensionality}
\newacronym{llpf}{LLPF}{local loss pre-filtering}
\newacronym{mitm}{MITM}{man-in-the-middle}
\newacronym{ae}{AE}{adversary entitie}
\newacronym{tof}{TOF}{time of fly}
\newacronym{rssi}{RSS}{received signal strength}
\newacronym{3d}{3D}{three dimensional}
\newacronym{aoa}{DoA}{Direction of Arrival}
\newacronym{sdp}{SDP}{semi-definite programming}
\newacronym{nlos}{NLOS}{Non-Line-of-Sight}
\newacronym{snr}{SNR}{Signal to Noise Ratio}
\newacronym{crb}{CRB}{Cramer-Rao bound}
\newacronym{lse}{LSE}{least squared estimation}
\newacronym{wlse}{WLSE}{weighted least squared estimation}
\newacronym{gd}{GD}{Gradient descend}
\newacronym{ap}{AP}{Access Points}
\newacronym{crlb}{CRLB}{Cramér-Rao Lower Bound}
\newacronym{tdoa}{TDoA}{Time Difference of Arrival}
\newacronym{sinr}{SINR}{Signal to Interference and Noise Ratio}
\newacronym{los}{LOS}{Line of Sight}
\newacronym{a2g}{A2G}{Air to Ground}
\newacronym{eu}{EU}{European Union}
\newacronym{umiav}{UMi-AV}{Urban Micro–Aerial Vehicle}
\newacronym{3gpp}{3GPP}{3rd Generation Partnership Project}
\newacronym{lae}{LAE}{Low Altitude Economy}
\newacronym{gnss}{GNSS}{Global Navigation Satellite System}
\newacronym{rf}{RF}{Radio Frequency}
\newacronym{gpdr}{GDPR}{General Data Protection Regulation}
\newacronym{5gnr}{5G-NR}{Fifth Generation New Radio}
\newacronym{otdoa}{OTDOA}{Observed Time Difference of Arrival}
\newacronym{prs}{PRS}{Positioning Reference Signals}
\newacronym{gnb}{gNB}{Next Generation Node B}
\newacronym{lmf}{LMF}{Localization Management Function}
\newacronym{amf}{AMF}{Access and Mobility Management Function}
\newacronym{lpp}{LPP}{LTE Positioning Protocol}
\newacronym{nrppa}{NRPPa}{NG-RAN Positioning Protocol A}
\newacronym{prc}{PRC}{Positioning Reference Configuration}
\newacronym{dlotdoa}{DL-OTDOA}{Downlink Observed Time Difference of Arrival}
\newacronym{ulotdoa}{UL-OTDOA}{Uplink Observed Time Difference of Arrival}
\newacronym{nas}{NAS}{Non-Access Stratum}
\newacronym{ngc}{NG-C}{Next Generation Control Plane}
\newacronym{tls}{TLS}{Transport Layer Security}
\newacronym{rsrp}{RSRP}{Reference Signal Received Power}
\newacronym{rof}{ROF}{RSS-based optimum finder}
\newacronym{tcv}{TCV}{Triangular Consistency Verification}
\newacronym{sdet}{SDET}{Static Distance-Error Thresholding}
\newacronym{rdef}{RDEF}{Recursive Distance-Error Filtering}
\newacronym{lawn}{LAWN}{Low-altitude Wireless Network}
\newacronym{ls}{LS}{Least Squares}
\newacronym{iot}{IoT}{Internet of Things}
\newacronym{iq}{IQIA-Net}{In-phase Quadrature Intra-attention Network}
\newacronym{qr}{QR}{Quality report}
\newacronym{ages}{AGES}{Adaptive Gain Exponential Smoother}
\newacronym{v2x}{V2X}{Vehicle-to-Everything}

\usepackage{tcolorbox}
\tcbuselibrary{many}

\newtcbtheorem[]{alg}{Algorithm}{fonttitle=\bfseries}{alg}

\usepackage[norelsize,linesnumbered,ruled]{algorithm2e}
\SetKwRepeat{Do}{do}{while}
\makeatletter
\newcommand{\removelatexerror} {\let\@latex@error\@gobble}
\makeatother

\begin{document}
	
	\title{Quality-Aware Denoising of Ultra-Short TDoA Measurements for 5G-NR UAV Localization}
	
	\author{
		\IEEEauthorblockN{Zexin~Fang\IEEEauthorrefmark{1},~Bin~Han\IEEEauthorrefmark{1},~Anjie~Qiu\IEEEauthorrefmark{1},~Zhuojun~Tian\IEEEauthorrefmark{2} and~Hans~D.~Schotten\IEEEauthorrefmark{1}\IEEEauthorrefmark{3}}
		\IEEEauthorblockA{
  \IEEEauthorrefmark{1}{RPTU University Kaiserslautern-Landau, Germany}; \IEEEauthorrefmark{2}{KTH Royal Institute of Technology, Sweden}\\ 
  \IEEEauthorrefmark{3}{German Research Center for Artificial Intelligence (DFKI), Germany.}}
	}
	
	\bstctlcite{IEEEexample:BSTcontrol}
	
	% make the title area
	\maketitle

	\begin{abstract}  
    Reliable positioning is essential for \glspl{uav} in safety-critical urban operations, yet achieving sub-meter accuracy under stringent latency constraints remains challenging. While \gls{3gpp} specifies repeated \gls{prs} transmissions for accurate \gls{tdoa} measurements, denoising techniques specifically tailored for extremely limited measurement sequences within \gls{3gpp} frameworks remain underexplored. We propose \gls{ages}, a lightweight filter combining exponentially weighted averaging with adaptive gains informed by \gls{3gpp} measurement quality reports. Simulations demonstrate \gls{ages} achieves $30-40\%$ reduction in positioning error with only $3-5$ repeated measurements while maintaining \gls{5gnr} infrastructure compatibility.
		
	\end{abstract}
    
	% Note that keywords are not normally used for peerreview papers.
	
	\IEEEpeerreviewmaketitle
	
	\glsresetall

	\section{Introduction}\label{sec:introduction}
 With the advancement of \gls{lae}, \glspl{uav} are increasingly deployed for safety-critical civil services such as emergency medical supply delivery, search and rescue operations, and infrastructure inspection in urban areas. As regulatory frameworks worldwide open airspace for commercial \gls{uav} operations, demand for robust positioning systems has intensified. While \gls{gnss} remains the primary positioning method, its vulnerability to signal degradation in urban canyons and multipath interference poses significant challenges for autonomous navigation. Research and standardization efforts have shifted toward terrestrial infrastructure-supported localization, with \gls{3gpp} progressively enhancing positioning capabilities: Release 16 introduced \gls{5gnr} positioning features including \gls{dlotdoa} and \gls{ulotdoa} with sub-meter accuracy targets, while Releases 17 and 18 targeted decimeter-level accuracy for industrial \gls{iot} and \gls{v2x} applications \cite{3gpp.38.305, 3enhancefre, 3gpp.38.859}. Commercial deployments are underway globally, including China Mobile and China Unicom in major cities, Verizon and AT\&T pilots in the United States, and Deutsche Telekom and Vodafone trials in Europe. Despite these advances, achieving reliable sub-meter accuracy in challenging propagation environments remains an open challenge for aerial platforms in dense urban settings.

To enhance localization accuracy in \gls{dlotdoa} positioning, \gls{prs} signals are transmitted periodically, enabling the \gls{ue} to obtain multiple \gls{tdoa} measurements over consecutive \gls{prs} occasions. These measurements, along with quality reports, are forwarded to the \gls{lmf} for temporal smoothing and position estimation. Emerging \gls{5gnr} positioning targets emphasize both low latency and high accuracy: Rel-17 defines a commercial positioning latency target of \emph{$\leq 100$ ms end-to-end}, with desirable latencies of \emph{10 ms} for industrial \gls{iot} use cases, alongside sub-meter accuracy requirements. However, meeting these dual objectives simultaneously presents a fundamental challenge: achieving the required latency necessitates processing measurements over ultra-short observation windows. Consequently, ultra-short measurement sequences are prevalent in practical \gls{5gnr} positioning for \glspl{uav} and industrial \gls{iot} devices. In dense urban environments, the \gls{ue} may only receive $3-7$ consecutive \gls{prs} measurements before reporting to the \gls{lmf} to satisfy real-time constraints. Standard smoothing and Kalman filtering methods typically assume longer measurement sequences. Applying conventional filters over short sequences either underperforms due to insufficient data for velocity modeling, or introduces delays that violate latency targets. As the result, there is a critical need to investigate lightweight denoising techniques specifically tailored to ultra-short sequences that:
\begin{enumerate*}[label=\emph{\roman*)}]
\item improve measurement reliability without extensive historical data,
\item maintain low computational complexity suitable for \gls{ue} or \gls{lmf} processing,
\item respect end-to-end latency constraints,
\item and provide robust performance with $3-7$ samples.
\end{enumerate*}
\begin{figure*}[t]
\centering
\includegraphics[width=0.70\linewidth]{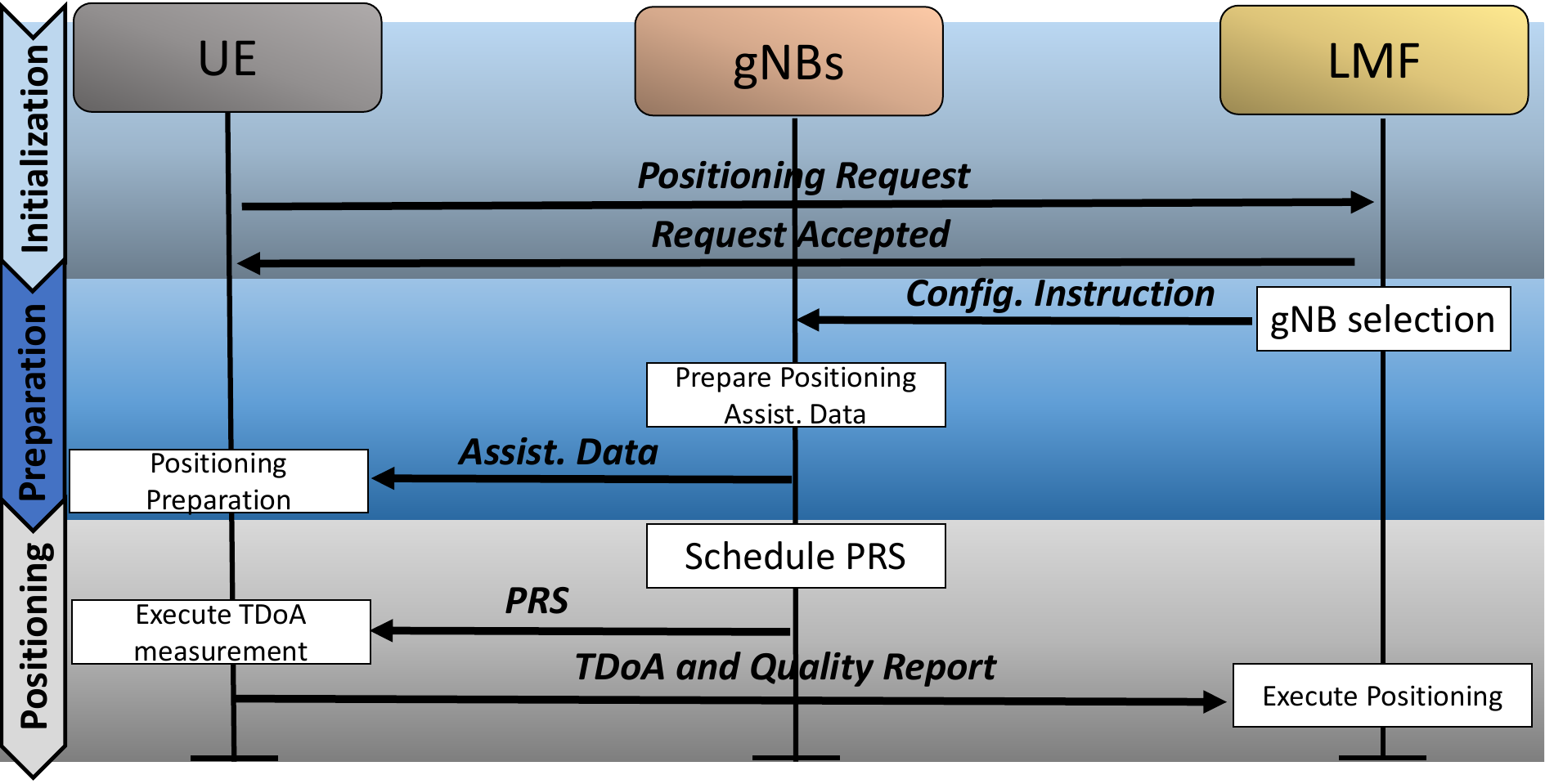}
\caption{\gls{5gnr} \gls{dlotdoa} localization procedure showing signaling flow between network entities.}
\label{fig:5gnrflow}
\end{figure*}

To the best of our knowledge, lightweight measurement-level filtering for ultra-short sequences in dynamic \gls{uav} scenarios remains largely unexplored. Several studies have investigated denoising techniques for limited observation windows. Palivonaite et al. propose algebraic short-term forecasting with mixed smoothing, demonstrating meaningful noise reduction with very short sequences \cite{Palivonaite2016}. Exponential smoothing methods have been widely studied for short-term smoothing and forecasting, emphasizing recent observations while requiring minimal historical data and computational complexity \cite{Bandara2020, Hyndman2008}. In acoustic source localization, Kalman filtering and recursive smoothing of \gls{tdoa} measurements have demonstrated effectiveness by directly processing noisy measurements \cite{Klee2006,8569848}. 

Inspired by these studies and tailored for \gls{5gnr}, we propose a Kalman-inspired filter using exponentially weighted averaging combined with Kalman-style gain-based updates and measurement quality reports defined in the \gls{3gpp} framework, detailed in Alg.~\ref{alg:spgd}. The rest of this paper is organized as follows: In Sec.\ref{sec:sys_model}, we introduce the relevant \gls{3gpp} framework and \gls{a2g} channel model. In Sec.\ref{sec:loctech}, we present the \gls{tdoa} measurement system model and proposed denoising techniques. In Sec.\ref{sec:eva}, we evaluate these techniques. Finally, we conclude in Sec.\ref{sec:con}.
\section{Preliminary}\label{sec:sys_model}

\subsection{\gls{3gpp} \gls{5gnr} framework}
We introduce a \gls{5gnr} localization system following the \gls{3gpp} architecture for \gls{otdoa}-based \gls{ue} positioning. The \gls{lmf} serves as the central positioning entity, coordinating localization procedures, computing \gls{ue} positions from reported measurements, and managing inter-cell synchronization for accurate multilateration. The \gls{amf} handles connection and mobility management, tunneling \gls{lpp} messages between the \gls{lmf} and \gls{ue}. The \glspl{gnb} transmit \gls{prs} and maintain precise time synchronization for \gls{tdoa}-based positioning.

The positioning procedure relies on three key protocols:
\begin{enumerate}
    \item \textbf{\gls{lpp}} facilitates signaling between the \gls{lmf} and \gls{ue} \cite{3gpp.37.355}. The \gls{lmf} sends measurement configuration specifying the \gls{prc}, instructing the \gls{ue} on which \gls{prs} to measure. The \gls{ue} performs \gls{tdoa} measurements and returns detailed reports to the \gls{lmf} for position computation.
    \item \textbf{\gls{nrppa}} enables the \gls{lmf} to coordinate \gls{prs} transmission across \glspl{gnb} \cite{3gpp.38.355}, specifying transmission timing, duration, frequency resources, and beam configuration for coordinated network deployment.
    \item \textbf{\gls{prs}} are downlink reference signals transmitted by \glspl{gnb} according to the configured \gls{prc}. The \gls{ue} measures \gls{tdoa} between signals from different \glspl{gnb} and reports measurements with \gls{qr} indicators. \gls{prs} can be time-multiplexed (avoiding inter-cell interference) or frequency-multiplexed (enabling simultaneous transmissions for reduced latency in mobile \gls{uav} scenarios).
\end{enumerate}

We employ \gls{dlotdoa}, where the \gls{ue} passively measures \gls{tdoa} and reports to the \gls{lmf} for position computation (Fig.~\ref{fig:5gnrflow}). This centralized approach minimizes signaling overhead and computational burden on \glspl{ue} while enabling sophisticated positioning algorithms. Specifically, we consider frequency-multiplexed \gls{prs}-enabled \gls{dlotdoa} as specified in Release 17 for high-accuracy, low-latency applications \cite{3enhancefre}.

   \subsection{\gls{a2g} channel model}
The \gls{a2g} propagation characteristics have been comprehensively analyzed by \gls{3gpp}, with corresponding models documented in \cite{3gpp-tr36.777}. For the \gls{umiav} scenario, \gls{los} probability is expressed as
\begin{equation}\label{eq:problos}
P_\text{los} =
\begin{cases}
1, & d_{\text{2D}} \le d_1,\\[4pt]
\left(1-\frac{d_1}{d_{\text{2D}}}\right)\exp\!\left(\frac{-d_{\text{2D}}}{p_1}\right)
+ \frac{d_1}{d_{\text{2D}}}, & d_{\text{2D}} > d_1,
\end{cases}
\end{equation}
where $d_{\text{2D}}$ represents the horizontal distance between the aerial vehicle and the terrestrial base station. The \gls{uav} altitude $h$ influences the model through parameters $d_1$ and $p_1$, defined as
\begin{equation}\label{eq:d1p1}
\begin{split}
d_1 &= \max\big(294.05\log_{10}(h) - 432.94,\; 18 \big),\\
p_1 &= 233.98\log_{10}(h) - 0.95.
\end{split}
\end{equation}
Combining \gls{los} and \gls{nlos} conditions, the average path-loss exponent $\eta$ (in dB), as defined in \cite{3gpp-tr36.777}, is given by
\begin{equation}
\begin{split}
\eta ={}& \big(4.32 - 0.76\log_{10}(h)\big)\left(1 - P_\text{los}\right) \\
&+ \big(2.225 - 0.05\log_{10}(h)\big)P_\text{los}.
\end{split}
\end{equation}
These expressions reveal that channel quality generally improves with increasing altitude due to higher \gls{los} probability, while degrading with increasing horizontal distance as \gls{nlos} conditions become more prevalent.
    \section{Methodology}\label{sec:loctech}
    \subsection{System model} 
  First, \gls{tdoa} measurements are derived through correlation of \glspl{prs}, which exhibit short pulse characteristics with desirable autocorrelation properties. Research in \cite{Venus2020ToA, Hechen2024ToA} demonstrates that dense multipath propagation fundamentally constrains \gls{tdoa} measurement accuracy via the Cramér-Rao lower bound $\sigma_d^2 \geq J_{\text{T}}^{-1}$, where the Fisher information matrix $J_{\text{T}}$ for \gls{tdoa} estimation is expressed as:
\begin{equation}\label{eq:toacrb}
J_{\text{T}} = {2 c^{-1} 4 \pi^2 \text{SINR}\gamma \beta^2 \sin^2(\phi)}.
\end{equation}
In this formulation, $c$ denotes the speed of light, $\beta$ the signal bandwidth, and $\text{SINR}$ quantifies the signal-to-interference-plus-noise ratio dominated by multipath effects and inter-cell interference. Given that \gls{3gpp} specifications enforce orthogonal \gls{prs} allocation across \glspl{gnb} and that \gls{a2g} propagation typically experiences limited multipath components, we apply the simplification $\text{SINR} \approx \text{SNR}$. The parameters $\gamma$ and $\sin^2(\phi)$ characterize whitening filter efficiency and degradation from path loss uncertainty, respectively. Analysis in \cite{WGKlaus2016} reveals that although both terms exhibit bandwidth dependence, $\sin^2(\phi)$ demonstrates significantly reduced sensitivity compared to $\gamma$, indicating that $\text{SNR}$, $\beta$, and $\gamma$ constitute the primary determinants of $J_{\text{T}}$. The whitening gain admits the approximation $\gamma \propto \ln(\beta)\beta^{-1}$ \cite{fang2025network}, yielding the simplified variance bound:
\begin{equation}\label{eq:toacrb_simplified}
\sigma_d^2 \propto {\frac{c}{8 \pi^2 \text{SNR}\beta\ln{\beta}} }.
\end{equation}
This relationship provides a theoretical foundation for weighted positioning algorithms to improve localization robustness.
\begin{figure}[t]
	\centering
	\includegraphics[width=0.82\linewidth]{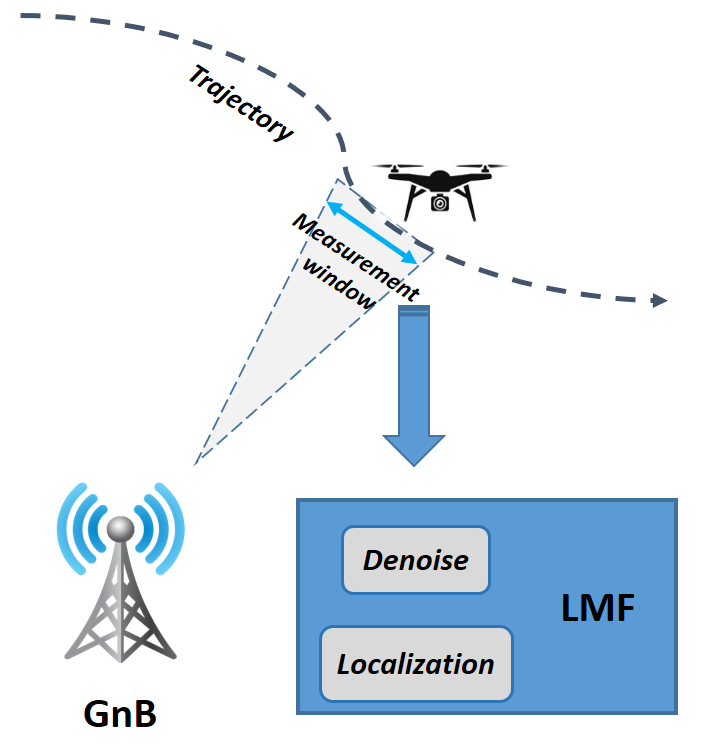}
	\caption{Illustration of measurement denoising window for ultra-short TDOA sequences}
	\label{fig:denoise_proccess}
\end{figure}

Second, achieving reliable position estimates necessitates repeated \gls{tdoa} measurements at the \gls{ue}, the proccess of \gls{tdoa} measurements from one \gls{gnb} and the denoising proccess is depicted in Fig.~\ref{fig:denoise_proccess}. We aggregate the ground-truth \gls{tdoa} observations across $N$ base stations over $K$ temporal instances into the measurement matrix $\bm{\tau}$:
\begin{equation}\nonumber
\bm{\tau} =
\begin{bmatrix}
\tau_1^1  & \tau_1^2 &\cdots &\tau_1^n & \cdots &\tau_1^N\\
\tau_2^1 & \tau_2^2 &\cdots &  \tau_2^n& \cdots &\tau_2^N\\
\vdots&\vdots&\ddots&\vdots&\ddots&\vdots\\
\tau_k^1 & \tau_k^2 &\cdots &  \tau_k^n& \cdots &\tau_k^N\\
\vdots&\vdots&\ddots&\vdots&\ddots&\vdots\\
\tau_K^1 & \tau_K^2 &\cdots &  \tau_K^n& \cdots &\tau_K^N
\end{bmatrix}.
\end{equation}
\gls{prs} transmissions occur at fixed periodicity $\Delta_t$. Since \gls{3gpp} specifications impose stringent synchronization requirements on \glspl{gnb}, and the \gls{lmf} manages synchronization errors, the inter-node timing offsets $\bm{\delta} = [\delta_1, \delta_2, \ldots, \delta_N]$ exhibit negligible drift over a short time span. The observed \gls{tdoa} matrix is therefore modeled as:
\begin{equation}
\bm{\tau}_R = \bm{\tau} + \bm{\delta} \times \mathbf{1}^{K \times 1} + \bm{\epsilon}_D,
\end{equation}
where $\bm{\epsilon}_D$ represents the measurement noise matrix. $\bm{\tau}_R$ can be then interpreted to the distances matrix $\bm{D} = \bm{\tau}_R\cdot c$, where $\mathbf{D} = \{ \tilde{d}_k^n \}_{n=1,k=1}^{N,K}$. One the distance obtained, localization can be performed by \gls{lmf}. 

    \subsection{Denoise techniques}\label{subsec:LSE}
    To address the challenge of ultra-short \gls{tdoa} sequences, we introduce several lightweight filtering techniques alongside our proposed approach:

    \textbf{Exponential Smoothing:} This method assigns exponentially decaying weights to historical measurements, emphasizing recent observations. The smoothed estimate is computed as $\hat{x}_k = \sum_{i=1}^{k} w_i x_i / \sum_{i=1}^{k} w_i$, where $w_i = \alpha^{k-i}$ and $\alpha \in (0,1)$ controls the decay rate. It requires minimal computational overhead and naturally adapts to time-varying signals.

   \textbf{Double Exponential Smoothing:} Extending simple exponential smoothing, DES incorporates trend estimation through two equations: level $\ell_k = \alpha x_k + (1-\alpha)(\ell_{k-1} + b_{k-1})$ and trend $b_k = \beta(\ell_k - \ell_{k-1}) + (1-\beta)b_{k-1}$. The smoothed output combines both as $\hat{x}_k = \ell_k + b_k$, enabling tracking of linear trends even with limited data.
   
   \textbf{Median Filter:} Computes the median value over a temporal window centered at each measurement point. This approach provides robustness against sporadic outliers by selecting the middle-ranked value within the window, though it does not account for velocity or motion trends in the data.

   \textbf{Savitzky-Golay Filter:} A polynomial-based smoothing approach that approximates the local signal behavior through least-squares polynomial fitting. By evaluating the fitted polynomial, this method attenuates measurement noise while preserving important signal characteristics such as peaks and trends.

   \textbf{Proposed \gls{ages}:} We propose an \gls{ages} specifically designed for \gls{tdoa} measurement denoising under ultra-short observation windows. Unlike classical Kalman filtering that requires explicit state-space models and process noise characterization, AGES integrates exponentially weighted moving averages with Kalman-style adaptive gain mechanism informed by measurement quality reports. \gls{ages} exploits \gls{3gpp} defined measurement reports to construct time-varying measurement covariances, enabling quality-aware filtering.
  
\begin{algorithm}[!ht]
  \caption{\gls{ages}}
  \label{alg:spgd}
   \DontPrintSemicolon
  Input: Distances measurements $\bm{d}$ and measurement quality reports; Forgetting factor $\alpha$.  \\
  Output: The latest distance measurement from all \glspl{gnb} $\mathcal{X}$. \\
  \SetKwProg{Fn}{Function}{ :}{end}
  \Fn{\emph{AGES}}{
     Extract \gls{snr} from measurement reports and combine them with the bandwidths of \gls{prs}, then convert to variance matrix $\mathbf{R}$ using Eq.~\ref{eq:toacrb_simplified}, where $\mathbf{R} = \{ r_k^n \}_{n=1,k=1}^{N,K}$. \\
    \For {$n = 1:N$ }
    {\For{$k = 1:K-1$}{$w_k = \alpha^{(K-2-k)}$\;
    $P_\text{pre}^n \mathrel{+}= {r_k^n}/{(K-1)}$\;
    $X_\text{pre}^n \mathrel{+}= {\tilde{d}^n_k w_k}/{(K-1)}$}
    $K^n_{\text{gain}} = P_{\text{pre}}^n/(P_{\text{pre}}^n+r_K^n)$\;
    $X_\text{cur}^n = X_\text{pre}^n + K^n_{\text{gain}}(\tilde{d}^n_K - X_\text{pre}^n)$
    }
    Output $\mathcal{X}= \{X_\text{cur}^n \mid 1\leq n\leq N\}$}
  \end{algorithm}
   
    \section{Evaluation of Methodology}\label{sec:eva}
\begin{figure}[!htpb]
		\centering
		\begin{subfigure}{.95\linewidth}
			\centering     
			\includegraphics[width=\linewidth]{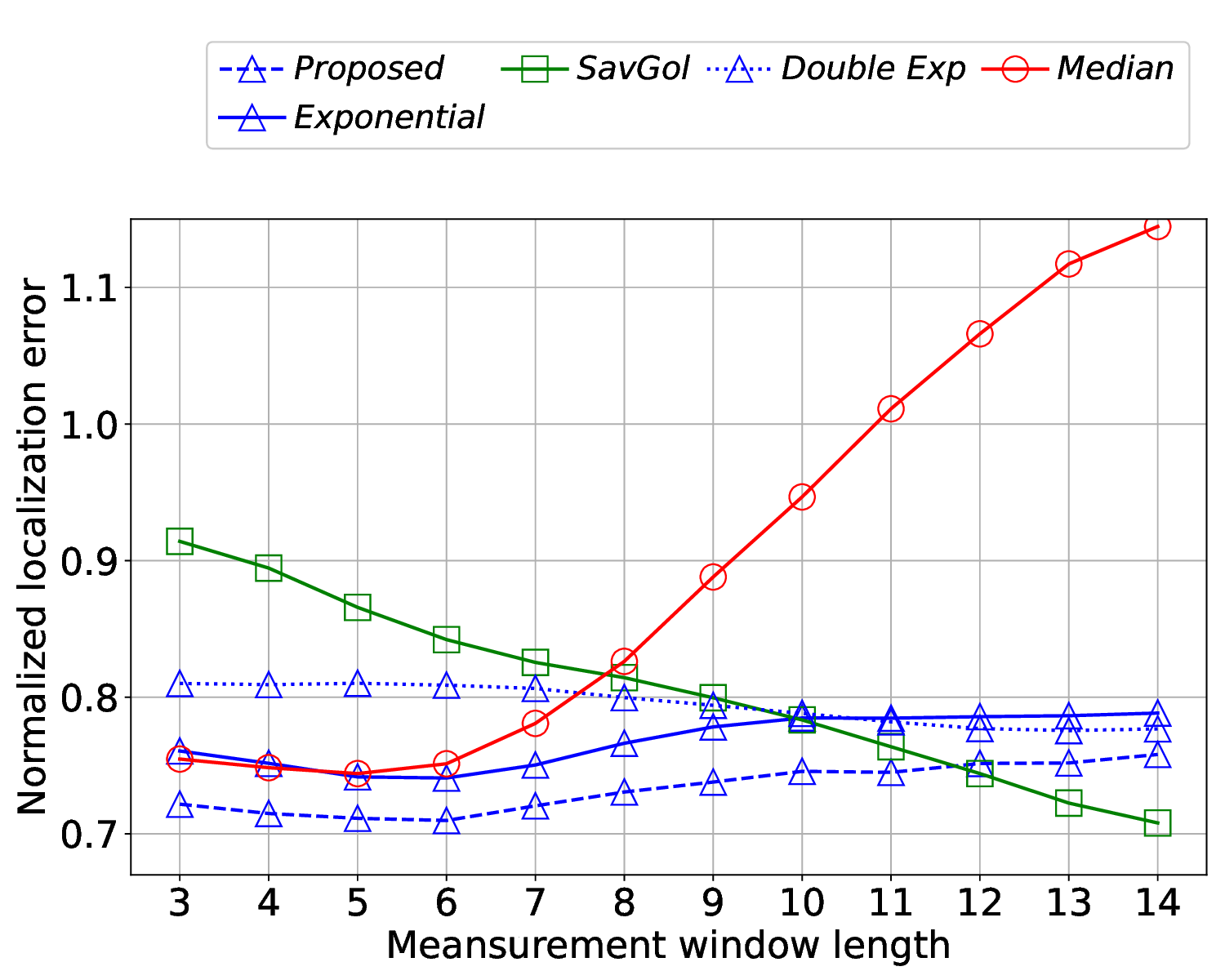}
			\subcaption{Altitude: $20$ m, speed: $90$ km/h}
			\label{subfig:lowalti}
		\end{subfigure}
%		\hfill
		\begin{subfigure}{.95\linewidth}
			\centering
			\includegraphics[width=\linewidth]{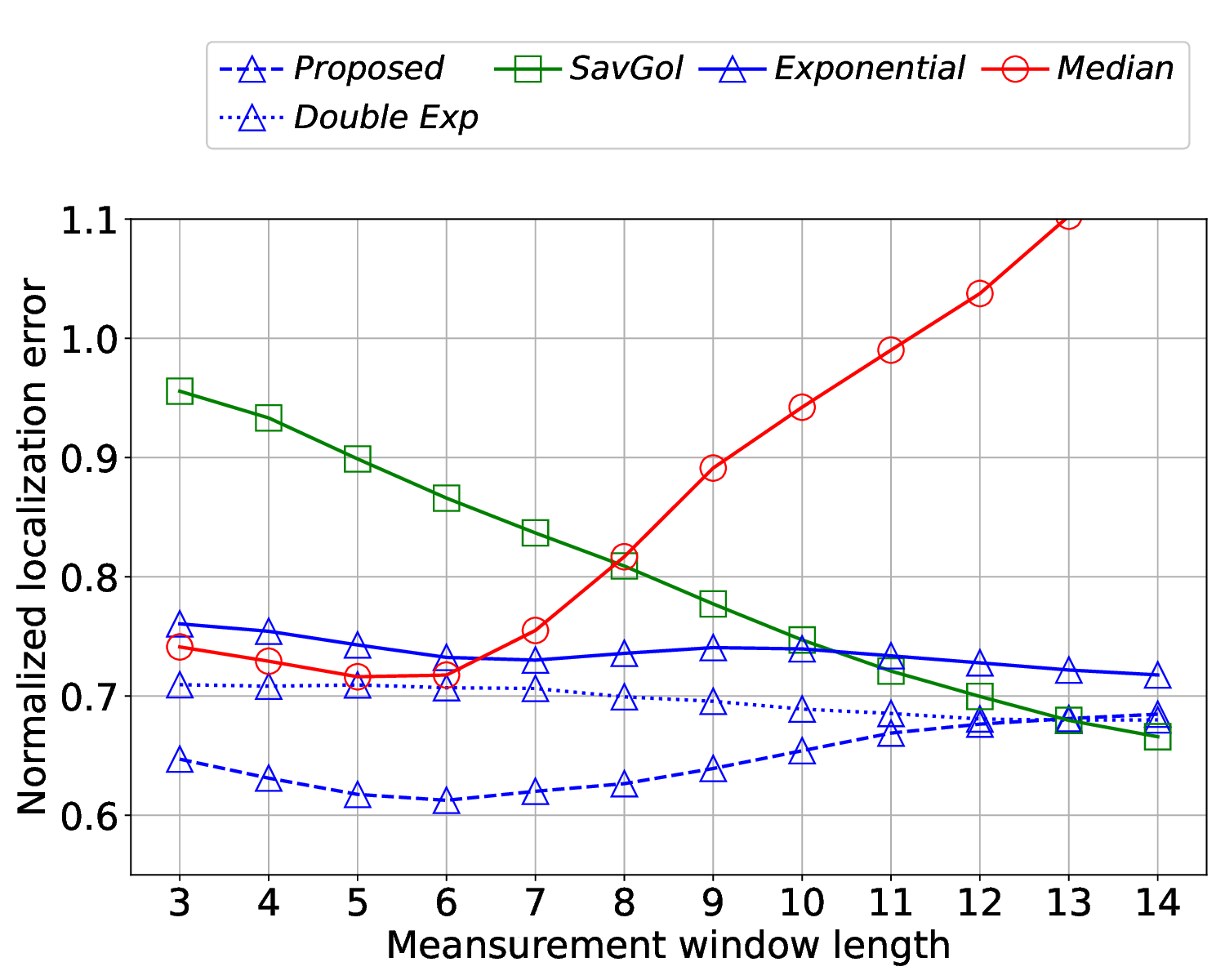}
			\subcaption{Altitude: $30$ m, speed: $90$ km/h}
			\label{subfig:normal}
		\end{subfigure}
        \begin{subfigure}{.95\linewidth}
			\centering
			\includegraphics[width=\linewidth]{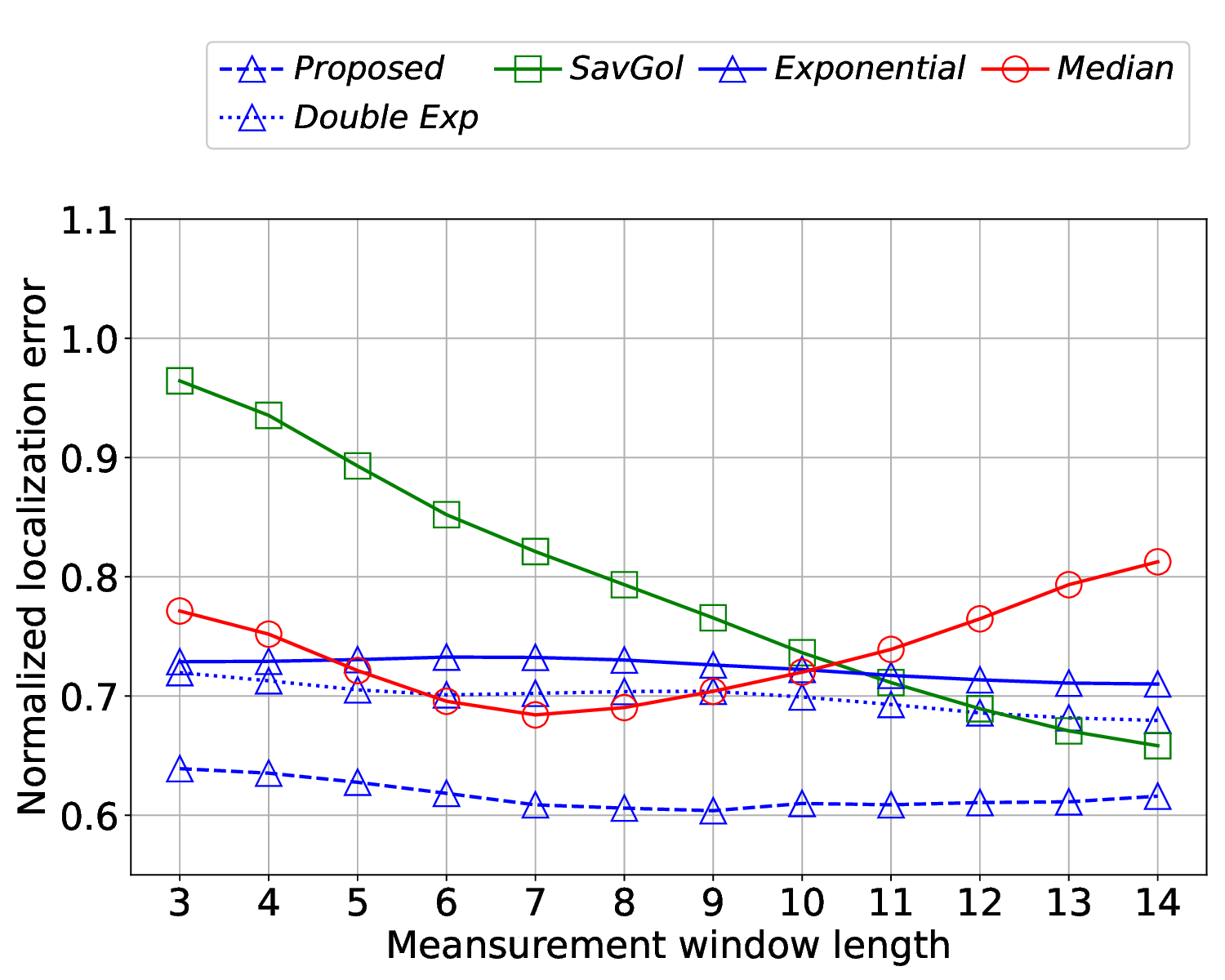}
			\subcaption{Altitude: $30$ m, speed: $50$ km/h}
			\label{subfig:slow}
		\end{subfigure}
     \caption{Performance of different denoising techniques. The normalized localization error is obtained by dividing by the localization error without denoising.}\label{fig:locprob}
	\end{figure}

    Next, we proceed to evaluate the aforementioned denoising techniques. We assume that the bilateration problem is resolved by the \gls{lmf}, and the \gls{uav} operates in a \gls{gnb}-dense urban environment. The \gls{uav} flies at approximately constant velocity within the measurement window, with minor speed fluctuations induced by mechanical dynamics and wind disturbances. The detailed simulation setup regarding \gls{tdoa} measurements and deployment is listed in Tab.~\ref{tab:setup1}. Additionally, the velocity jitter is modeled as $10\%$ of the nominal velocity with random directional perturbations. The simulation results are consolidated in Fig.~\ref{fig:locprob}, with each data point averaged over $1000$ Monte Carlo runs. It is worth noting that, due to latency requirements, practical \gls{prs} measurements are typically limited to only a few frames. However, we extend the measurement window beyond $10$ frames to provide deeper insights into the behavior and limitations of the aforementioned denoising techniques. Subsequently, the denoised measurements combined with \gls{gnb} coordinates are used to localize the \gls{uav} using the efficient gradient descent algorithm introduced in \cite{GD2025fang,vtcfang2023,fangvtcfall2024}.

   \begin{table}[!t]
		\centering
		\caption{Simulation setup 1}
		\label{tab:setup1}
		\begin{tabular}{>{}m{0.2cm} | m{1.6cm} l m{3.7cm}}
			\toprule[2px]
%			\rowcolor{white}
			&\textbf{Parameter}&\textbf{Value}&\textbf{Remark}\\
			\midrule[1px]        
			&$f_c$&$3.5$ GHz& Carrier frequency\\

			&$K$&$(0.1,3.0)$& Rician factors\\
			&$N_p$&$4$& Number of multipath\\
            & $\tau_\text{max}$ & 2e-7 s &  Maximum delay spread\\ 
            & $P_\text{t}$ & 15 dBm &  Transmitting power\\ 
            & $N_\text{o}$ & -91 dBm &  Noise floor\\ 
            & $\beta_n$ & 10 Mhz &  Bandwidth\\ 
            & $\sigma_t$ & $1 \mu s $&  Average synchronization error\\
            \midrule[1px]
            \multirow{-9.9}{*}{\rotatebox{90}{\textbf{TDOA}}}
            & $h_u $&$[20,30]$ m& \gls{uav} altitude\\ 
            &$R$&$120$ m & Node coverage\\
			\multirow{-1.7}{*}{\rotatebox{90}{\textbf{Deploy.}}}&$h_n$&$\sim\mathcal{U}(0,5)$ m& Node coverage \\
            &$V_u$&$[50, 90]$ km/h& \gls{uav} velocity\\
            &$\Delta_t$&$20$ ms& \gls{prs} interval\\
            &$N$&$8$ ms& Assigned \gls{gnb} number\\
            \bottomrule[2px]
		\end{tabular}
	\end{table}
   
  In Fig.~\ref{subfig:lowalti}, the \gls{uav} cruises at relatively low altitude, where \gls{a2g} channels typically exhibit degraded quality due to increased \gls{nlos} probability. \gls{ages} achieves the best performance for short measurement windows. The median filter demonstrates effective denoising only when the measurement window is small, as it completely disregards velocity consideration. When the window is sufficiently short, \gls{uav} displacement remains minimal, making the constant-value assumption approximately valid. However, as the measurement window expands, median filter performance deteriorates significantly due to its inability to track motion-induced trends. Similarly, \gls{ages} exhibits slightly reduced denoising effectiveness with longer windows, as it does not explicitly model velocity dynamics. The Savitzky-Golay filter shows performance degradation as the measurement window increases, though it could potentially surpass our method with substantially longer windows at the cost of prohibitive latency. Double exponential smoothing underperforms compared to \gls{ages} and simple exponential smoothing, as the limited number of measurements introduces substantial estimation error when attempting to model both level and trend components. In Fig.~\ref{subfig:normal}, the \gls{uav} operates at higher altitude, where measurement quality improves substantially due to enhanced channel conditions and reduced \gls{nlos} components. Under these favorable propagation conditions, double exponential smoothing performance notably improves, as higher-quality measurements enable more reliable trend estimation with reduced noise interference. In Fig.~\ref{subfig:slow}, the \gls{uav} travels at reduced velocity, significantly altering the filtering performance characteristics. The performance degradation of the median filter is substantially mitigated, as slower motion results in smaller \gls{tdoa} variations across the measurement window, better aligning with its implicit stationarity assumption. Notably, the optimal measurement window length extends to approximately $9$ frames, demonstrating the method's adaptiveness to velocity changes.

    \section{Conclusion}\label{sec:con}
This work established that effective \gls{tdoa} denoising for \gls{uav} positioning is achievable with severely constrained measurement sequences. By exploiting standardized \gls{3gpp} measurement reports, significant noise reduction is obtained without additional signaling overhead or complex motion models. Our evaluation revealed that velocity-agnostic methods like median filtering fail under motion, while trend-based approaches like double exponential smoothing require more data than available. \gls{ages} addresses this through adaptive weighting that accounts for measurement recency and quality. This work provides a practical solution compatible with existing \gls{5gnr} infrastructure, enabling reliable autonomous \gls{uav} navigation in dense urban environments.
     \section*{Acknowledgment}
	This work is supported by the Federal Ministry of Research, Technology and Space of Germany via the project
Open6GHub+ (16KIS2406). B. Han (bin.han@rptu.de) is the
corresponding author. 
    
\bibliographystyle{IEEEtran}
\bibliography{references}

\end{document}